\documentclass[11pt]{article}
\usepackage[dvips]{graphicx}
\usepackage{epsfig}
\usepackage{amssymb}
\usepackage{color}
\usepackage{amsmath}
\usepackage{cite}
\usepackage[left]{lineno}
\begin{document}
\centerline{\LARGE EUROPEAN ORGANIZATION FOR NUCLEAR RESEARCH}
%
%
\vspace{10mm} {\flushright{
CERN-PH-EP-2015-093 \\
2 April 2015\\
Revised version:\\30 April 2015\\
}}
\vspace{-35mm}
%
%

\vspace{40mm}

\vspace{10mm}

\begin{center}
\boldmath
{\bf {\Large \boldmath{Search for the dark photon in $\pi^0$ decays}}}
\unboldmath
\end{center}
\begin{center}
{\Large The NA48/2 collaboration}\\
\end{center}

\begin{abstract}
A sample of $1.69\times 10^7$ fully reconstructed $\pi^0\to\gamma e^+e^-$ decay candidates collected by the NA48/2 experiment at CERN in 2003--2004 is analysed to search for the dark photon ($A'$) production in the $\pi^0\to\gamma A'$ decay followed by the prompt $A'\to e^+e^-$ decay. No signal is observed, and an exclusion region in the plane of the dark photon mass $m_{A'}$ and mixing parameter $\varepsilon^2$ is established. The obtained upper limits on $\varepsilon^2$ are more stringent than the previous limits in the mass range $9~{\rm MeV}/c^2<m_{A'}<70~{\rm MeV}/c^2$. The NA48/2 sensitivity to the dark photon production in the $K^\pm\to\pi^\pm A'$ decay is also evaluated.
\end{abstract}

\begin{center}
{\it Accepted for publication in Physics Letters B}
\end{center}

\newpage
\begin{center}
{\Large The NA48/2 Collaboration}\\
\vspace{2mm}
 J.R.~Batley,
 G.~Kalmus,
 C.~Lazzeroni$\,$\footnotemark[1]$^,$\footnotemark[2],
 D.J.~Munday,
 M.W.~Slater$\,$\footnotemark[1],
 S.A.~Wotton \\
{\em \small Cavendish Laboratory, University of Cambridge,
Cambridge, CB3 0HE, UK$\,$\footnotemark[3]} \\[0.2cm]
 R.~Arcidiacono$\,$\footnotemark[4],
 G.~Bocquet,
 N.~Cabibbo$\,$\renewcommand{\thefootnote}{\fnsymbol{footnote}}%
\footnotemark[2]\renewcommand{\thefootnote}{\arabic{footnote}},
 A.~Ceccucci,
 D.~Cundy$\,$\footnotemark[5],
 V.~Falaleev,
 M.~Fidecaro,
 L.~Gatignon,
 A.~Gonidec,
 W.~Kubischta,
 A.~Norton$\,$\footnotemark[6],
 A.~Maier,\\
 M.~Patel$\,$\footnotemark[7],
 A.~Peters\\
{\em \small CERN, CH-1211 Gen\`eve 23, Switzerland} \\[0.2cm]
 S.~Balev$\,$\renewcommand{\thefootnote}{\fnsymbol{footnote}}%
\footnotemark[2]\renewcommand{\thefootnote}{\arabic{footnote}},
 P.L.~Frabetti,
 E.~Gersabeck$\,$\footnotemark[8],
 E.~Goudzovski$\,$\renewcommand{\thefootnote}{\fnsymbol{footnote}}%
\footnotemark[1]\renewcommand{\thefootnote}{\arabic{footnote}}$^,$\footnotemark[1]$^,$\footnotemark[2]$^,$\footnotemark[9],
 P.~Hristov$\,$\footnotemark[10],
 V.~Kekelidze,
 V.~Kozhuharov$\,$\footnotemark[11]$^,$\footnotemark[12],
 L.~Litov$\,$\footnotemark[11],
 D.~Madigozhin,
 N.~Molokanova,
 I.~Polenkevich,
 Yu.~Potrebenikov,
 S.~Stoynev$\,$\footnotemark[13],
 A.~Zinchenko \\
{\em \small Joint Institute for Nuclear Research, 141980 Dubna (MO), Russia} \\[0.2cm]
 E.~Monnier$\,$\footnotemark[14],
 E.~Swallow,
 R.~Winston$\,$\footnotemark[15]\\
{\em \small The Enrico Fermi Institute, The University of Chicago,
Chicago, IL 60126, USA}\\[0.2cm]
 P.~Rubin$\,$\footnotemark[16],
 A.~Walker \\
{\em \small Department of Physics and Astronomy, University of
Edinburgh, Edinburgh, EH9 3JZ, UK} \\[0.2cm]
 W.~Baldini,
 A.~Cotta Ramusino,
 P.~Dalpiaz,
 C.~Damiani,
 M.~Fiorini,
 A.~Gianoli, \\
 M.~Martini,
 F.~Petrucci,
 M.~Savri\'e,
 M.~Scarpa,
 H.~Wahl \\
 {\em \small Dipartimento di Fisica e Scienze della Terra dell'Universit\`a e Sezione
dell'INFN di Ferrara, \\ I-44122 Ferrara, Italy} \\[0.2cm]
 A.~Bizzeti$\,$\footnotemark[17],
 M.~Lenti,
 M.~Veltri$\,$\footnotemark[18] \\
{\em \small Sezione dell'INFN di Firenze, I-50019 Sesto Fiorentino, Italy} \\[0.2cm]
 M.~Calvetti,
 E.~Celeghini,
 E.~Iacopini,
 G.~Ruggiero$\,$\footnotemark[10] \\
{\em \small Dipartimento di Fisica dell'Universit\`a e Sezione
dell'INFN di Firenze, I-50019 Sesto Fiorentino, Italy} \\[0.2cm]
 M.~Behler,
 K.~Eppard,
 K.~Kleinknecht,
 P.~Marouelli,
 L.~Masetti,
 U.~Moosbrugger,\\
 C.~Morales Morales$\,$\footnotemark[19],
 B.~Renk,
 M.~Wache,
 R.~Wanke,
 A.~Winhart$\,$\footnotemark[1]\\
{\em \small Institut f\"ur Physik, Universit\"at Mainz, D-55099 Mainz, Germany$\,$\footnotemark[20]} \\[0.2cm]
 D.~Coward$\,$\footnotemark[21],
 A.~Dabrowski$\,$\footnotemark[10],
 T.~Fonseca Martin,
 M.~Shieh,
 M.~Szleper,\\
 M.~Velasco,
 M.D.~Wood$\,$\footnotemark[21] \\
{\em \small Department of Physics and Astronomy, Northwestern
University, Evanston, IL 60208, USA}\\[0.2cm]
 P.~Cenci,
 M.~Pepe,
 M.C.~Petrucci \\
{\em \small Sezione dell'INFN di Perugia, I-06100 Perugia, Italy} \\[0.2cm]
 G.~Anzivino,
 E.~Imbergamo,
 A.~Nappi$\,$\renewcommand{\thefootnote}{\fnsymbol{footnote}}%
\footnotemark[2]\renewcommand{\thefootnote}{\arabic{footnote}},
 M.~Piccini,
 M.~Raggi$\,$\footnotemark[12],
 M.~Valdata-Nappi \\
{\em \small Dipartimento di Fisica dell'Universit\`a e
Sezione dell'INFN di Perugia, I-06100 Perugia, Italy} \\[0.2cm]
 C.~Cerri,
 R.~Fantechi \\
{\em Sezione dell'INFN di Pisa, I-56100 Pisa, Italy} \\[0.2cm]
 G.~Collazuol$\,$\footnotemark[22],
 L.~DiLella,
 G.~Lamanna$\,$\footnotemark[12],
 I.~Mannelli,
 A.~Michetti \\
{\em Scuola Normale Superiore e Sezione dell'INFN di Pisa, I-56100
Pisa, Italy} \\[0.2cm]
 F.~Costantini,
 N.~Doble,
 L.~Fiorini$\,$\footnotemark[23],
 S.~Giudici,
 G.~Pierazzini$\,$\renewcommand{\thefootnote}{\fnsymbol{footnote}}%
\footnotemark[2]\renewcommand{\thefootnote}{\arabic{footnote}},
 M.~Sozzi,
 S.~Venditti$\,$\footnotemark[10]\\
{\em Dipartimento di Fisica dell'Universit\`a e Sezione dell'INFN di
Pisa, I-56100 Pisa, Italy} \\[0.2cm]
\newpage
 B.~Bloch-Devaux$\,$\footnotemark[24],
 C.~Cheshkov$\,$\footnotemark[25],
 J.B.~Ch\`eze,
 M.~De Beer,
 J.~Derr\'e,
 G.~Marel,
 E.~Mazzucato,
 B.~Peyaud,
 B.~Vallage \\
{\em \small DSM/IRFU -- CEA Saclay, F-91191 Gif-sur-Yvette, France} \\[0.2cm]
 M.~Holder,
 M.~Ziolkowski \\
{\em \small Fachbereich Physik, Universit\"at Siegen, D-57068 Siegen, Germany$\,$\footnotemark[26]} \\[0.2cm]
 C.~Biino,
 N.~Cartiglia,
 F.~Marchetto \\
{\em \small Sezione dell'INFN di Torino, I-10125 Torino, Italy} \\[0.2cm]
 S.~Bifani$\,$\footnotemark[1],
 M.~Clemencic$\,$\footnotemark[10],
 S.~Goy Lopez$\,$\footnotemark[27]\\
{\em \small Dipartimento di Fisica dell'Universit\`a e
Sezione dell'INFN di Torino, I-10125 Torino, Italy} \\[0.2cm]
 H.~Dibon,
 M.~Jeitler,
 M.~Markytan,
 I.~Mikulec,
 G.~Neuhofer,
 L.~Widhalm$\,$\renewcommand{\thefootnote}{\fnsymbol{footnote}}%
\footnotemark[2]\renewcommand{\thefootnote}{\arabic{footnote}} \\
{\em \small \"Osterreichische Akademie der Wissenschaften, Institut
f\"ur Hochenergiephysik,\\ A-10560 Wien, Austria$\,$\footnotemark[28]} \\[0.5cm]
\end{center}

%
\renewcommand{\thefootnote}{\fnsymbol{footnote}}
\footnotetext[1]{Corresponding author, email: eg@hep.ph.bham.ac.uk}
\footnotetext[2]{Deceased}
\renewcommand{\thefootnote}{\arabic{footnote}}
\footnotetext[1]{Now at: School of Physics and Astronomy, University of Birmingham, Birmingham, B15 2TT, UK}
\footnotetext[2]{Supported by a Royal Society University Research Fellowship (UF100308, UF0758946)}
\footnotetext[3]{Funded by the UK Particle Physics and Astronomy Research Council, grant PPA/G/O/1999/00559}
\footnotetext[4]{Now at: Universit\`a degli Studi del Piemonte Orientale e Sezione dell'INFN di Torino, I-10125 Torino, Italy}
\footnotetext[5]{Now at: Istituto di Cosmogeofisica del CNR di Torino, I-10133 Torino, Italy}
\footnotetext[6]{Now at: Dipartimento di Fisica e Scienze della Terra dell'Universit\`a e Sezione dell'INFN di Ferrara, I-44122 Ferrara, Italy}
\footnotetext[7]{Now at: Department of Physics, Imperial College, London, SW7 2BW, UK}
\footnotetext[8]{Now at: Physikalisches Institut, Ruprecht-Karls-Universit\"at Heidelberg, D-69120 Heidelberg, Germany}
\footnotetext[9]{Supported by ERC Starting Grant 336581}
\footnotetext[10]{Now at: CERN, CH-1211 Gen\`eve 23, Switzerland}
\footnotetext[11]{Now at: Faculty of Physics, University of Sofia ``St. Kl. Ohridski'', 1164 Sofia, Bulgaria, funded by the Bulgarian National Science Fund under contract DID02-22}
\footnotetext[12]{Now at: Laboratori Nazionali di Frascati, I-00044 Frascati, Italy}
\footnotetext[13]{Now at: Northwestern University, Evanston, IL 60208, USA}
\footnotetext[14]{Now at: Centre de Physique des Particules de Marseille, IN2P3-CNRS, Universit\'e de la M\'editerran\'ee, F-13288 Marseille, France}
\footnotetext[15]{Now at: School of Natural Sciences, University of California, Merced, CA 95344, USA}
\footnotetext[16]{Now at: School of Physics, Astronomy and Computational Sciences, George Mason
University, Fairfax, VA 22030, USA}
\footnotetext[17]{Also at Dipartimento di Scienze Fisiche, Informatiche e Matematiche, Universit\`a di Modena e Reggio Emilia, I-41125 Modena, Italy}
\footnotetext[18]{Also at Istituto di Fisica, Universit\`a di Urbino, I-61029 Urbino, Italy}
\footnotetext[19]{Present address: Helmholtz-Institut Mainz, Universit\"at Mainz, D-55099 Mainz, Germany}
\footnotetext[20]{Funded by the German Federal Minister for Education and Research under contract 05HK1UM1/1}
\footnotetext[21]{Now at: SLAC, Stanford University, Menlo Park, CA 94025, USA}
\footnotetext[22]{Now at: Dipartimento di Fisica dell'Universit\`a e Sezione dell'INFN di Padova, I-35131 Padova, Italy}
\footnotetext[23]{Now at: Instituto de F\'isica Corpuscular IFIC, Universitat de Valencia, E-46071 Valencia, Spain}
\footnotetext[24]{Now at: Dipartimento di Fisica dell'Universit\`a di Torino, I-10125 Torino, Italy}
\footnotetext[25]{Now at: Institut de Physique Nucl\'eaire de Lyon, Universit\'e Lyon I, F-69622 Villeurbanne, France}
\footnotetext[26]{Funded by the German Federal Minister for Research and Technology (BMBF) under contract 056SI74}
\footnotetext[27]{Now at: Centro de Investigaciones Energeticas
Medioambientales y Tecnologicas, E-28040 Madrid, Spain}
\footnotetext[28]{Funded by the Austrian Ministry for Traffic and Research under the contract GZ 616.360/2-IV GZ 616.363/2-VIII, and by the Fonds f\"ur Wissenschaft und Forschung FWF Nr.~P08929-PHY}

\newpage


\section*{Introduction}

Kaons are a source of tagged neutral pion decays, and high intensity kaon experiments provide opportunities for precision $\pi^0$ decay measurements. The NA48/2 experiment at the CERN SPS collected a large sample of charged kaon ($K^\pm$) decays in flight, corresponding to about $2\times 10^{11}$ $K^\pm$ decays in the fiducial decay volume. This letter reports the search for a hypothetical dark photon (DP, denoted $A'$) using a large sample of tagged $\pi^0$ mesons from identified $K^\pm\to\pi^\pm\pi^0$ and $K^\pm\to\pi^0\mu^\pm\nu$ decays.

In a rather general set of hidden sector models with an extra $U(1)$ gauge symmetry~\cite{ho86}, the interaction of the DP with the visible sector proceeds through kinetic mixing with the Standard Model (SM) hypercharge. Such scenarios with GeV-scale dark matter provide possible explanations to the observed rise in the cosmic-ray positron fraction with energy and the muon gyromagnetic ratio $(g-2)$ measurement~\cite{po09}. The DP is characterized by two a priori unknown parameters, the mass $m_{A'}$ and the mixing parameter $\varepsilon^2$. Its possible production in the $\pi^0$ decay and its subsequent decay proceed via the chain $\pi^0\to\gamma A'$, $A'\to e^+e^-$. The expected branching fraction of the above $\pi^0$ decay is~\cite{batell09}
\begin{equation}
{\cal B}(\pi^0\to\gamma A') = 2\varepsilon^2 \left(1-\frac{m_{A'}^2}{m_{\pi^0}^2}\right)^3 {\cal B}(\pi^0\to\gamma\gamma),
\label{eq:br}
\end{equation}
which is kinematically suppressed as $m_{A'}$ approaches $m_{\pi^0}$. In the DP mass range $2m_e<m_{A'}<m_{\pi^0}$ accessible in pion decays, the only allowed tree-level decay into SM fermions is $A'\to e^+e^-$, while the loop-induced SM decays ($A'\to 3\gamma$, $A'\to\nu\bar\nu$) are highly suppressed. Therefore, for a DP decaying only into SM particles, ${\cal B}(A'\to e^+e^-)\approx 1$, and the expected total decay width is~\cite{batell09}
\begin{equation}
\Gamma_{A'} \approx \Gamma(A'\to e^+e^-) = \frac{1}{3} \alpha\varepsilon^2 m_{A'} \sqrt{1-\frac{4m_e^2}{m_{A'}^2}}\left(1+\frac{2m_e^2}{m_{A'}^2}\right).
\end{equation}
It follows that, for $2m_e\ll m_{A'}<m_{\pi^0}$, the DP mean proper lifetime $\tau_{A'}$ satisfies the relation
\begin{equation}
c\tau_{A'} = \hbar c / \Gamma_{A'} \approx 0.8~{\mu\rm m} \times \left(\frac{10^{-6}}{\varepsilon^2}\right) \times \left(\frac{100~{\rm MeV}/c^2}{m_{A'}}\right).
\end{equation}
This analysis is performed assuming that the DP decays at the production point ({\it prompt decay}), which is valid for sufficiently large values of $m_{A'}$ and $\varepsilon^2$, as quantified in Section~\ref{sec:dp}. In this case, the DP production and decay signature is identical to that of the Dalitz decay $\pi^0_D\to e^+e^-\gamma$, which therefore represents an irreducible but well controlled background and determines the sensitivity.

The NA48/2 experiment provides pure $\pi^0_D$ decay samples through the reconstruction of $K^\pm\to\pi^\pm\pi^0$ and $K^\pm\to\pi^0\mu^\pm\nu$ decays (denoted $K_{2\pi}$ and $K_{\mu 3}$). Additionally, the $K^\pm\to\pi^\pm\pi^0\pi^0$ decay (denoted $K_{3\pi}$) is considered as a background in the $K_{\mu 3}$ sample. The $K^\pm\to\pi^0 e^\pm\nu$ decay is not considered for this analysis because of the ambiguity due to three $e^\pm$ particles in the final state.

\section{Beam, detector and data sample}
\label{sec:experiment}

The NA48/2 experiment used simultaneous $K^+$ and $K^-$ beams produced by 400~GeV/$c$ primary CERN SPS protons impinging on a beryllium target. Charged particles with momenta of $(60\pm3)$ GeV/$c$ were selected by an achromatic system of four dipole magnets which split the two beams in the vertical plane and recombined them on a common axis. The beams then passed through collimators and a series of quadrupole magnets, and entered a 114~m long cylindrical vacuum tank with a diameter of 1.92 to 2.4~m containing the fiducial decay region. Both beams had an angular divergence of about 0.05~mrad, a transverse size of about 1~cm, and were aligned with the longitudinal axis of the detector within 1~mm.

The vacuum tank was followed by a magnetic spectrometer housed in a vessel filled with helium at nearly atmospheric pressure, separated from the vacuum by a thin ($0.3\%~X_0$) $\rm{Kevlar}\textsuperscript{\textregistered}$ window.
An aluminium beam pipe of 158~mm outer diameter traversing the centre of the spectrometer (and all the following detectors) allowed the undecayed beam particles to continue their path in vacuum. The spectrometer consisted of four drift chambers (DCH) with an octagonal transverse width of 2.9~m: DCH1, DCH2 located upstream and DCH3, DCH4 downstream of a dipole magnet that provided a horizontal transverse momentum kick of 120~MeV/$c$ for charged particles. Each DCH was composed of eight planes of sense wires. The DCH space point resolution was 90~$\mu$m in both horizontal and vertical directions, and the momentum resolution was $\sigma_p/p = (1.02 \oplus 0.044\cdot p)\%$, with $p$ expressed in GeV/$c$. The spectrometer was followed by a plastic scintillator hodoscope (HOD) with a transverse size of about 2.4 m, consisting of a plane of vertical and a plane of horizontal strip-shaped counters arranged in four quadrants (each logically divided into four regions). The HOD provided time measurements of charged particles with 150~ps resolution. It was followed by a liquid krypton electromagnetic calorimeter (LKr), an almost homogeneous ionization chamber with an active volume of 7 m$^3$ of liquid krypton, $27~X_0$ deep, segmented transversally into 13248 projective $\sim\!2\!\times\!2$~cm$^2$ cells. The LKr energy resolution was $\sigma_E/E=(3.2/\sqrt{E}\oplus9/E\oplus0.42)\%$, the spatial resolution for an isolated electromagnetic shower was $(4.2/\sqrt{E}\oplus0.6)$~mm in both horizontal and vertical directions, and the time resolution was
$2.5~{\rm ns}/\sqrt{E}$, with $E$ expressed in GeV. The LKr was followed by a hadronic calorimeter and a muon detector, both not used in the present analysis. A detailed description of the beamline and detector can be found in Ref.~\cite{fa07,ba07}.

The NA48/2 experiment collected data in 2003--2004, during about 100 days of efficient data taking in total. A two-level trigger chain was employed to collect $K^\pm$ decays with at least three charged tracks in the final state~\cite{ba07}. At the first level (L1), a coincidence of hits in the two planes of the HOD was required to occur in at least two of 16 non-overlapping regions. The second level (L2) performed online reconstruction of trajectories and momenta of charged particles based on the DCH information. The L2 logic was based on the multiplicities and kinematics of reconstructed tracks and two-track vertices.

A GEANT3-based~\cite{geant} Monte Carlo (MC) simulation including full beamline, detector geometry and material description, magnetic fields, local inefficiencies, misalignment and their time variations throughout the running period is used to evaluate the detector response.

\boldmath
\section{Simulation of the $\pi^0_D$ background}
\unboldmath
\label{sec:background}

Simulations of the $K_{2\pi}$, $K_{\mu 3}$ and $K_{3\pi}$ decays followed by the $\pi^0_D$ decay (denoted $K_{2\pi D}$, $K_{\mu3 D}$ and $K_{3\pi D}$) are performed to evaluate the integrated kaon flux and to estimate the irreducible $\pi^0_D$ background to the DP signal. The $K_{2\pi}$ and $K_{\mu 3}$ decays are simulated including final-state radiation~\cite{ga06}. The $\pi^0_D$ decay is simulated using the lowest-order differential decay rate~\cite{mi72}
\begin{equation}
\label{eq:dgdxdy}
\frac{d^2\Gamma}{dxdy} = \Gamma_0\frac{\alpha}{\pi}|F(x)|^2\frac{(1-x)^3}{4x}\left(1+y^2+\frac{r^2}{x}\right),
\end{equation}
where $\Gamma_0$ is the $\pi^0\to\gamma\gamma$ decay rate, $r = 2m_e/m_{\pi^0}$, and $F(x)$ is the pion transition form factor (TFF). The kinematic variables are
\begin{equation}
x = \frac{(Q_1+Q_2)^2}{m_{\pi^0}^2} = (m_{ee}/m_{\pi^0})^2, ~~~~~ y = \frac{2P(Q_1-Q_2)}{m_{\pi^0}^2(1-x)},
\end{equation}
where $Q_1$, $Q_2$ and $P$ are the four-momenta of the two electrons ($e^\pm$) and the pion ($\pi^0$), and $m_{ee}$ is the invariant mass of the $e^+e^-$ pair.

Radiative corrections to the $\pi^0_D$ decay are implemented following the approach of Mikaelian and Smith~\cite{mi72} revised recently to provide an improved numerical precision~\cite{hu15}: the differential decay rate is modified by a radiative correction factor that depends on $x$ and $y$. In this approach, inner bremsstrahlung photon emission is not simulated, and its effects on the acceptance are not taken into account.

The TFF is conventionally parameterized as $F(x)=1+ax$. Vector meson dominance models expect the slope parameter to be $a\approx (m_{\pi^0}/m_\rho)^2 \approx 0.03$~\cite{la85}, while detailed calculations based on dispersion theory obtain $a = 0.0307\pm0.0006$~\cite{ho14}. Experimentally, the PDG average value $a=0.032\pm0.004$~\cite{pdg} is dominated by an $e^+e^-\to e^+e^-\pi^0$ measurement in the space-like region~\cite{cello}, while the most accurate measurements from $\pi^0$ decays have an uncertainty of 0.03. The precision on the radiative corrections to the $\pi^0_D$ decay is limited: in particular, the missing correction to the measured TFF slope due to two-photon exchange is estimated to be $\Delta a=+0.005$~\cite{ka06}. Therefore the background description cannot rely on the precise inputs from either experiment or theory.

An effective TFF slope is obtained from a fit to the measured $m_{ee}$ spectrum itself to provide a satisfactory background description (as quantified by a $\chi^2$ test) in the kinematic range $m_{ee}>8~{\rm MeV}/c^2$. The low $m_{ee}$ region is not considered for the DP search as the acceptance computation is less robust due to the steeply falling geometric acceptance at low $m_{ee}$ and lower electron identification efficiency at low momentum.


\section{Event reconstruction and selection}
\label{sec:selection}


Event selections for the $K_{2\pi}$ and $K_{\mu3}$ decays followed by the prompt $\pi^0\to\gamma A'$, $A'\to e^+e^-$ decay chain are employed. These two selections are identical up to the momentum, invariant mass and particle identification conditions. The principal selection criteria are listed below.
\begin{itemize}
\item Three-track vertices are reconstructed by extrapolation of track segments from the upstream part of the spectrometer into the decay volume, taking into account the measured Earth's magnetic field, stray fields due to magnetization of the vacuum tank, and multiple scattering.
\item The presence of a three-track vertex formed by a pion ($\pi^\pm$) or muon ($\mu^\pm$) candidate and two opposite sign electron ($e^\pm$) candidates is required. Particle identification is based on energy deposition in the LKr calorimeter ($E$) and momentum measured by the spectrometer ($p$). Pions from $K_{2\pi}$ decays and muons from $K_{\mu3}$ decays are kinematically constrained to the momentum range above 5~GeV/$c$, while the momentum spectra of electrons originating from $\pi^0$ decays are soft, peaking at 3~GeV/$c$. Therefore, $p>5~{\rm GeV}/c$ and $E/p<0.85$ ($E/p<0.4$) are required for the pion (muon) candidate, while $p>2.75~{\rm GeV}/c$ and $(E/p)_{\rm min}<E/p<1.15$, where $(E/p)_{\rm min}=0.80$ for $p<5~{\rm GeV}/c$ and $(E/p)_{\rm min}=0.85$ otherwise, are required for the electron candidates. The lower momentum cut and the weaker $E/p$ cut for low momentum electrons are optimised to compensate for the degraded energy resolution (as quantified in Section~\ref{sec:experiment}). The electron identification inefficiency decreases with momentum and does not exceed 0.5\% in the signal momentum range, while the muon identification inefficiency is below 0.1\%. The pion identification inefficiency varies between 1\% and 2\% depending on momentum, and is applied to the simulation using measurements from data samples of fully reconstructed $K_{2\pi}$ and $K^\pm\to 3\pi^\pm$ decays.
\item The tracks forming the vertex are required to be in the fiducial geometric acceptances of the DCH, HOD and LKr detectors. Track separations in the DCH1 plane should exceed 2~cm to reject photon conversions, and electron track separations from electron (pion, muon) tracks in the LKr front plane should exceed 10~cm (25~cm) to minimize the effects of shower overlap.
\item A single isolated LKr energy deposition cluster is considered as a photon candidate. It should be compatible in time with the tracks, and separated by at least 10~cm (25~cm) from the electron (pion, muon) impact points. The reconstructed photon energy should be above 3~GeV to reduce the effects of non-linearity (which is about 1\% at 3~GeV energy) and degraded resolution at low energy.
\item An event is classified as a $K_{2\pi}$ or $K_{\mu3}$ candidate based on the presence of a pion or a muon candidate and the following criteria. The total reconstructed momentum of the three tracks and the photon candidate should be in the range from 53 to 67~GeV/$c$ (below 62~GeV/$c$) for the $K_{2\pi}$ ($K_{\mu 3}$) candidates. The squared total reconstructed transverse momentum with respect to the nominal beam axis ($p_T^2$) should be below $5\times 10^{-4}$~$({\rm GeV}/c)^2$ for the $K_{2\pi}$ candidates, and in the range from $5\times 10^{-4}$ to 0.04~$({\rm GeV}/c)^2$ for the $K_{\mu 3}$ candidates. The two $p_T^2$ intervals do not overlap, therefore the $K_{2\pi}$ and $K_{\mu3}$ event selections are mutually exclusive.
\item The reconstructed invariant mass of the $e^+e^-\gamma$ system is required to be compatible with the nominal $\pi^0$ mass $m_{\pi^0}$~\cite{pdg}: $|m_{ee\gamma}-m_{\pi^0}|<8~{\rm MeV}/c^2$. This interval corresponds to $\pm 5$ times the resolution on $m_{ee\gamma}$.
\item For the $K_{2\pi}$ selection, the reconstructed invariant mass of the $\pi^\pm e^+e^-\gamma$ system should be compatible with the nominal $K^\pm$ mass~\cite{pdg}: $474~{\rm MeV}/c^2<m_{\pi ee\gamma}<514~{\rm MeV}/c^2$. For the $K_{\mu 3}$ selection, the squared missing mass $m_{\rm miss}^2=(P_K-P_\mu-P_{\pi^0})^2$, where $P_\mu$ and $P_{\pi^0}$ are the reconstructed $\mu^\pm$ and $\pi^0$ four-momenta, and $P_K$ is the nominal kaon four-momentum, should be compatible to the missing neutrino mass: $|m_{\rm miss}^2|<0.01~{\rm GeV}^2/c^4$. The resolutions on $m_{\pi ee\gamma}$ and $m_{\rm miss}^2$ are 4.0~MeV/$c^2$ and $1.6\times 10^{-3}~{\rm GeV}^2/c^4$, respectively.
\item The DP mass cut: $|m_{ee}-m_{A'}|<\Delta m(m_{A'})$, where $m_{A'}$ is the assumed DP mass, and $\Delta m(m_{A'})$ is the half-width of the DP search window depending on $m_{A'}$ defined in Section~\ref{sec:dp}.
\end{itemize}

In addition to the above {\it individual DP selections} for the $K_{2\pi}$ and $K_{\mu3}$ decays, the {\it joint DP selection} is also considered: an event passes the joint selection if it passes either the $K_{2\pi}$ or the $K_{\mu3}$ selection. The acceptance of the joint selection $A_{\rm DP}$ for any process is equal to the sum of acceptances of the two mutually exclusive individual selections. Additionally, the {\it Dalitz decay} selections for the $K_{2\pi D}$ and $K_{\mu3D}$ decays are considered: they differ from the DP selections by the absence of the DP mass cut.


\begin{table}[p]
\begin{center}
\caption{Numbers of data events passing the $K_{2\pi D}$ and $K_{\mu 3D}$ selections, and  acceptances of these selections evaluated with MC simulations. The statistical errors on the acceptances are negligible.\label{tab:sel}}
\vspace{2mm}
\begin{tabular}{lcc}
\hline
& $K_{2\pi D}$ selection& $K_{\mu 3D}$ selection \\ 
\hline
\rule{0pt}{11pt}%
Data candidates: & $N_{2\pi D} = 1.38\times 10^7$ & $N_{\mu 3D} = 0.31\times 10^7$ \\
\hline
Acceptances:\\
for $K_{2\pi D}$ decay & $A_\pi(K_{2\pi D})=3.71\%$ & $A_\mu(K_{2\pi D})=0.11\%$ \\ 
for $K_{\mu 3D}$ decay & $A_\pi(K_{\mu3 D})=0.03\%$ & $A_\mu(K_{\mu3 D})=4.17\%$ \\ 
for $K_{3\pi D}$ decay & $A_\pi(K_{3\pi D})=0\phantom{.03\%}$ & $A_\mu(K_{3\pi D})=0.06\%$ \\ 
\hline
\end{tabular}
\end{center}
\end{table}

\begin{figure}[p]
\begin{center}
\resizebox{0.5\textwidth}{!}{\includegraphics{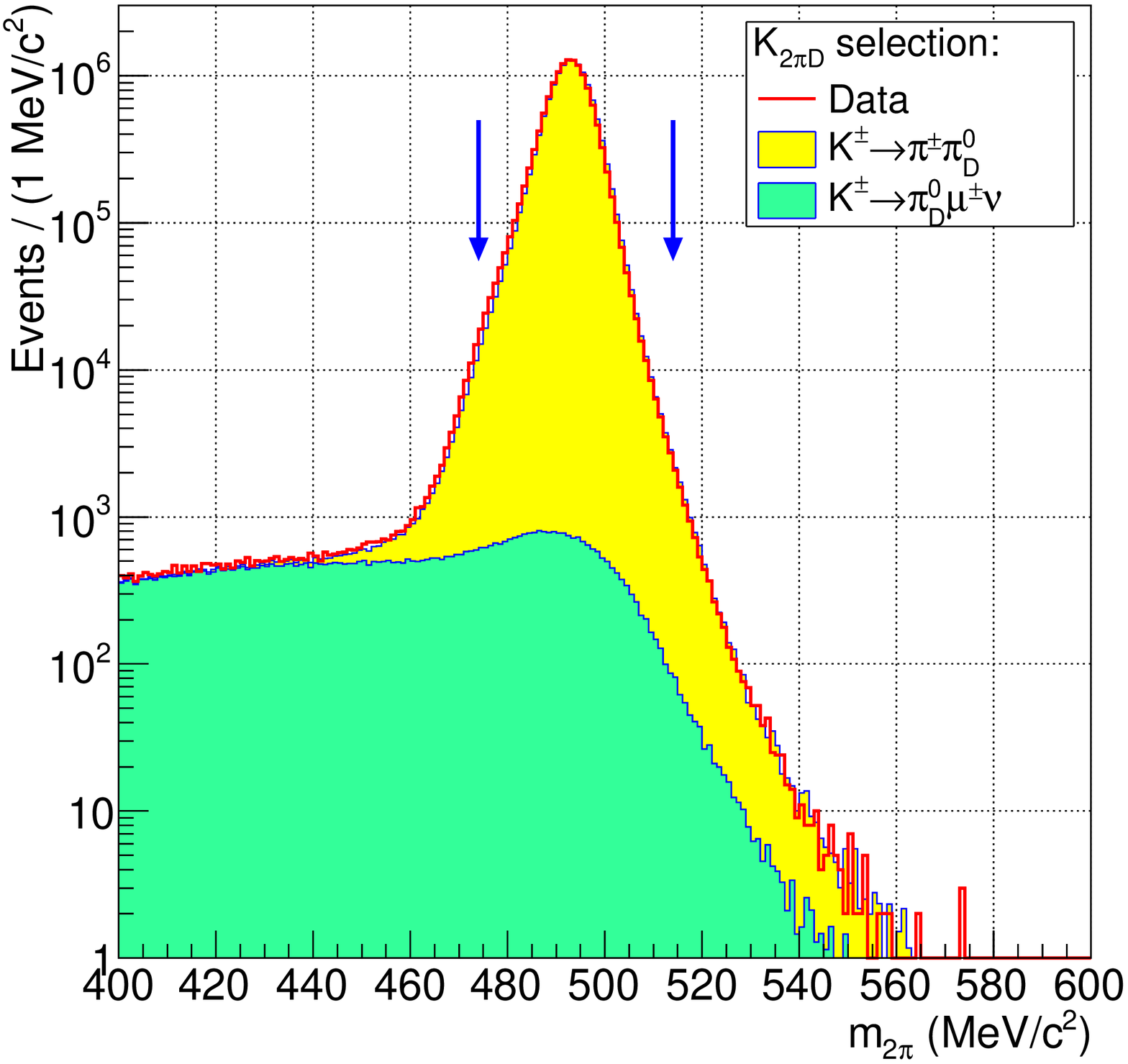}}%
\resizebox{0.5\textwidth}{!}{\includegraphics{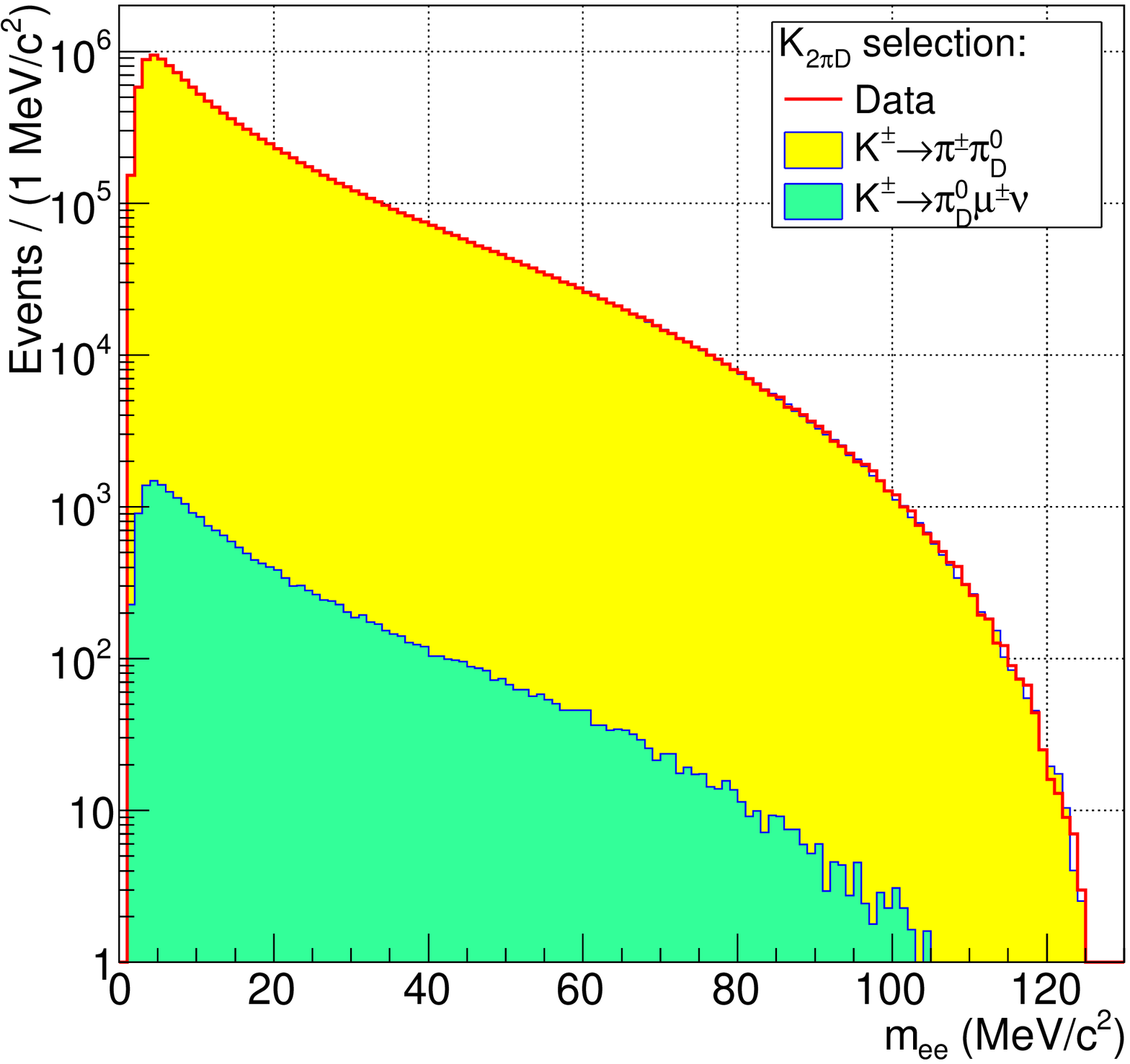}}\\
\resizebox{0.5\textwidth}{!}{\includegraphics{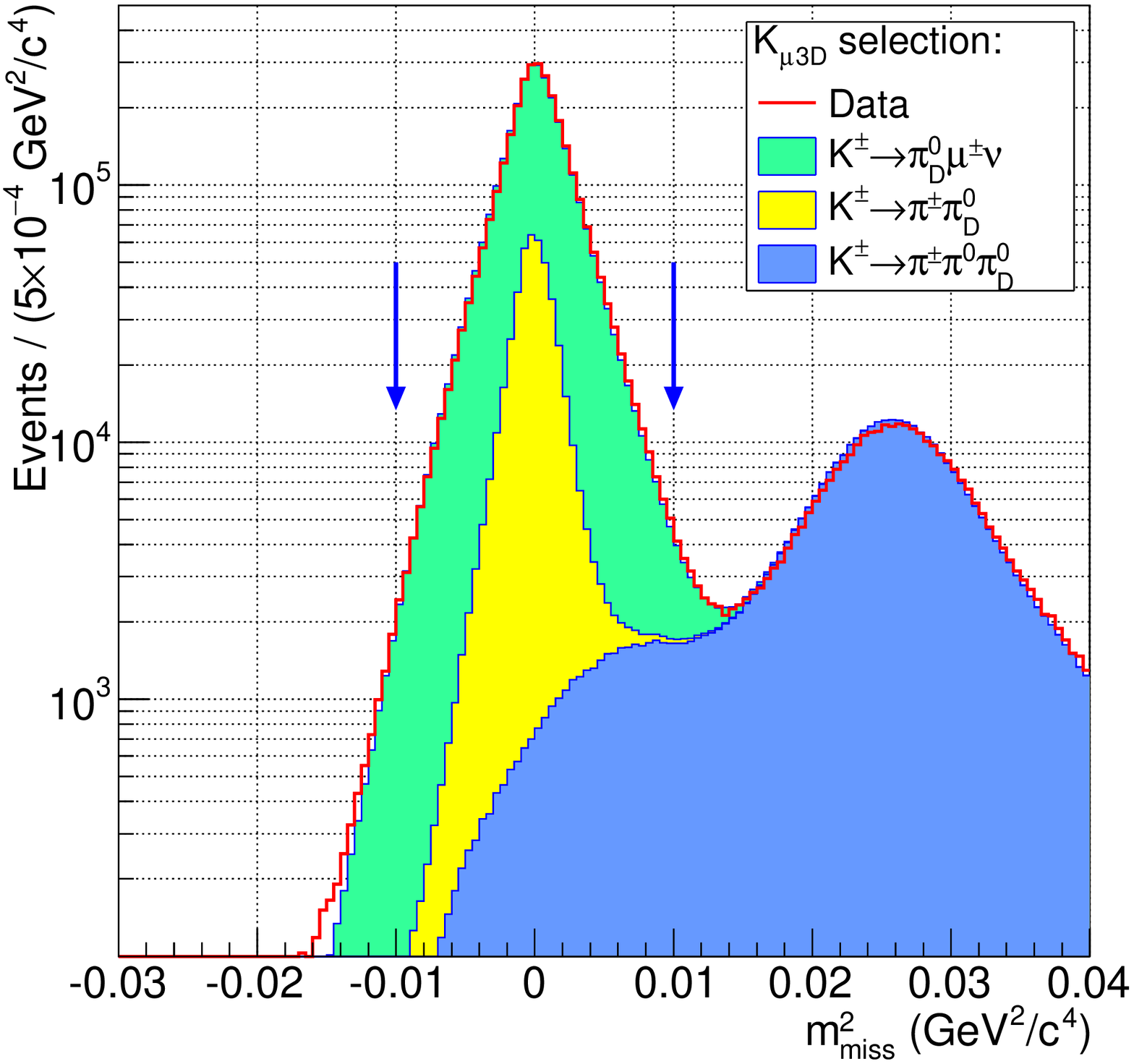}}%
\resizebox{0.5\textwidth}{!}{\includegraphics{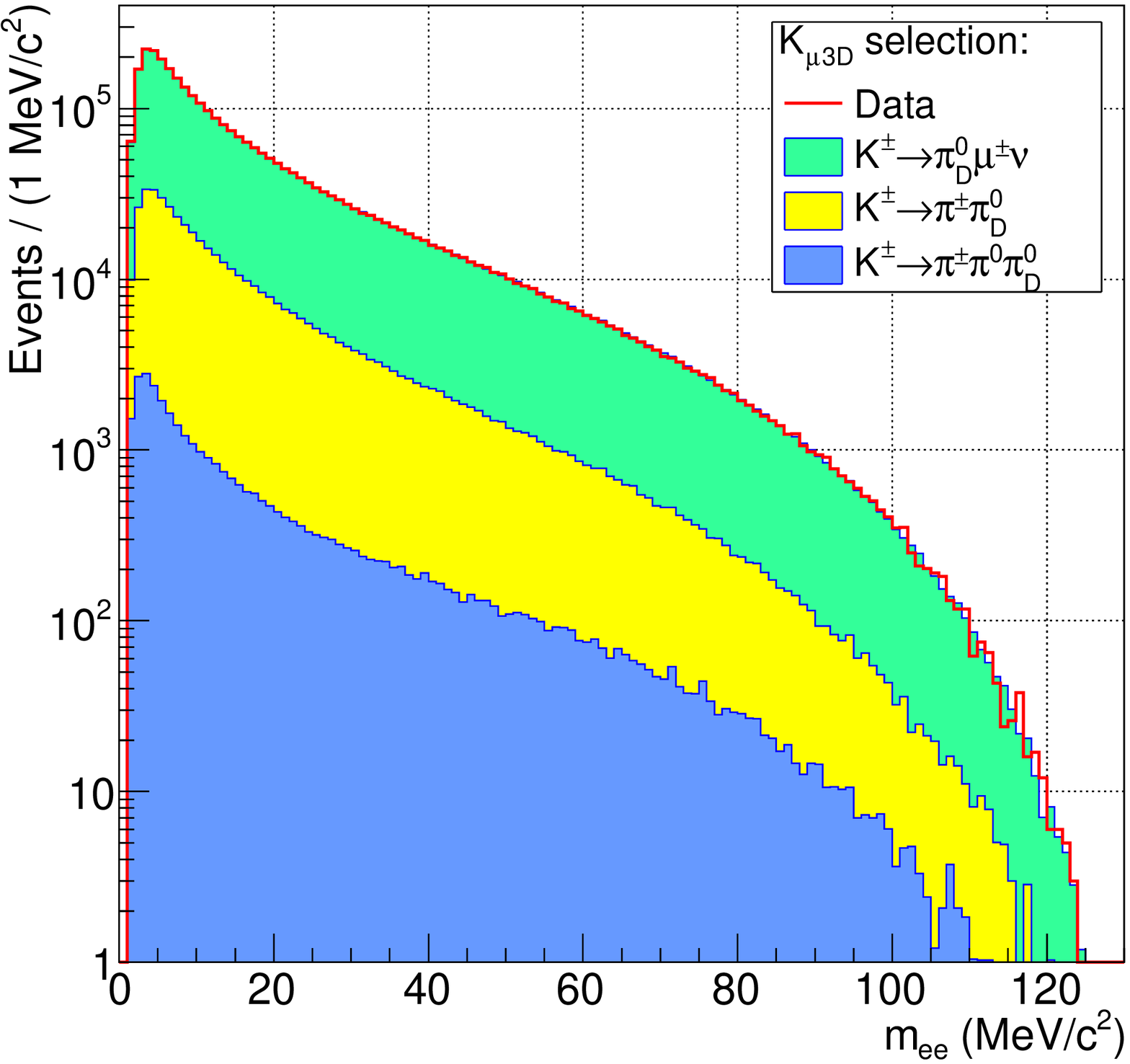}}
\end{center}
\vspace{-15mm}
\caption{Invariant mass distributions of data and MC events passing the $K_{2\pi D}$ (top row) and
$K_{\mu3 D}$ (bottom row) selections. The signal mass regions are indicated with vertical arrows. A dark photon signal would correspond to a spike in the $m_{ee}$ distributions (right column).}
\label{fig:mass}
\end{figure}

\boldmath
\section{Integrated kaon flux and $\pi^0_D$ data sample}
\unboldmath
\label{sec:flux}

The number of $K^\pm$ decays in the 98~m long fiducial decay region is computed as
\begin{equation}
N_K = \frac{N_{2\pi D}}{\left[{\cal B}(K_{2\pi})A_\pi(K_{2\pi D}) + {\cal B}(K_{\mu3})A_\pi(K_{\mu3D})\right] {\cal B}(\pi^0_D) e_1 e_2} =
(1.57\pm0.05)\times 10^{11},
\end{equation}
where $N_{2\pi D}$ is the number of data candidates reconstructed within the $K_{2\pi D}$ selection, $A_\pi(K_{2\pi D})$ and $A_\pi(K_{\mu 3D})$ are the acceptances of the $K_{2\pi D}$ selection for the $K_{2\pi D}$ and $K_{\mu 3D}$ decays evaluated with MC simulations, ${\cal B}(K_{2\pi})$, ${\cal B}(K_{\mu3})$, ${\cal B}(\pi^0_D)$ are the nominal branching fractions of the involved decay modes~\cite{pdg}, and $e_1=(99.75\pm 0.01)\%$, $e_2=(97.50\pm 0.04)\%$ are the efficiencies of the L1 and L2 trigger algorithms measured from downscaled control samples collected simultaneously with the main data set. A similar but statistically less precise value of $N_K$ is obtained from the number of data events passing the $K_{\mu 3D}$ selection, and the corresponding acceptances $A_\mu$ and trigger efficiencies. All numerical quantities are summarized in Table~\ref{tab:sel}. The number of $\pi^0_D$ candidates reconstructed with the joint Dalitz decay selection is $1.69\times 10^7$. The uncertainty on $N_K$ is dominated by the limited precision on ${\cal B}(\pi^0_D)$.

The analysis takes into account the cross-feeding between decay modes. In particular, $A_\pi(K_{\mu 3D})/[(A_\pi(K_{\mu 3D})+A_\mu(K_{\mu 3D})]=0.7\%$ of the reconstructed $K_{\mu 3D}$ events are classified as $K_{2\pi D}$ due to the low neutrino momentum. Conversely, about 3\% of the reconstructed $K_{2\pi D}$ events are classified as $K_{\mu 3D}$ due to $\pi^\pm\to\mu^\pm\nu$ decays in flight. $K_{3\pi D}$ decays constitute about 1\% of the $K_{\mu 3D}$ candidates.

The reconstructed invariant mass spectra ($m_{\pi ee\gamma}$, $m_{\rm miss}^2$ and $m_{ee}$) of data and MC events passing the Dalitz decay selections, with the MC samples normalised to the data using the estimated value of $N_K$, are shown in Fig.~\ref{fig:mass}.


\section{Search for the dark photon signal}
\label{sec:dp}

A scan for a DP signal in the mass range $9~{\rm MeV}/c^2 \le m_{A'} < 120~{\rm MeV}/c^2$ is performed. The lower boundary of the mass range is determined by the limited accuracy of the $\pi^0_D$ background simulation at low $e^+e^-$ mass (Section~\ref{sec:background}). At high DP mass approaching the upper limit of the mass range, the sensitivity to the mixing parameter $\varepsilon^2$ is not competitive with the existing limits due to the kinematic suppression of the $\pi^0\to\gamma A'$ decay.

The resolution on $m_{ee}$ as a function of $m_{ee}$ evaluated with MC simulation is parameterized as $\sigma_{m}(m_{ee}) = 0.067~{\rm MeV}/c^2 + 0.0105\cdot m_{ee}$, and varies from $0.16~{\rm MeV}/c^2$ to $1.33~{\rm MeV}/c^2$ over the mass range of the scan. The intrinsic DP width $\Gamma_{A'}$ is negligible with respect to $\sigma_{m}$. The mass step of the scan and the half-width of the DP search window are defined, depending on the value of $A'$ mass, as $\sigma_{m}(m_{A'})/2$ and $\Delta m=1.5\sigma_m(m_{A'})$, respectively (and both are rounded to the nearest multiple of 0.02 MeV/$c^2$). The search window width has been optimised with MC simulations to achieve the highest expected sensitivity to the DP signal, determined by a trade-off between $\pi^0_D$ background fluctuation and signal acceptance. In total, 404 DP mass values are tested.

\begin{figure}[p]
\begin{center}
\resizebox{0.5\textwidth}{!}{\includegraphics{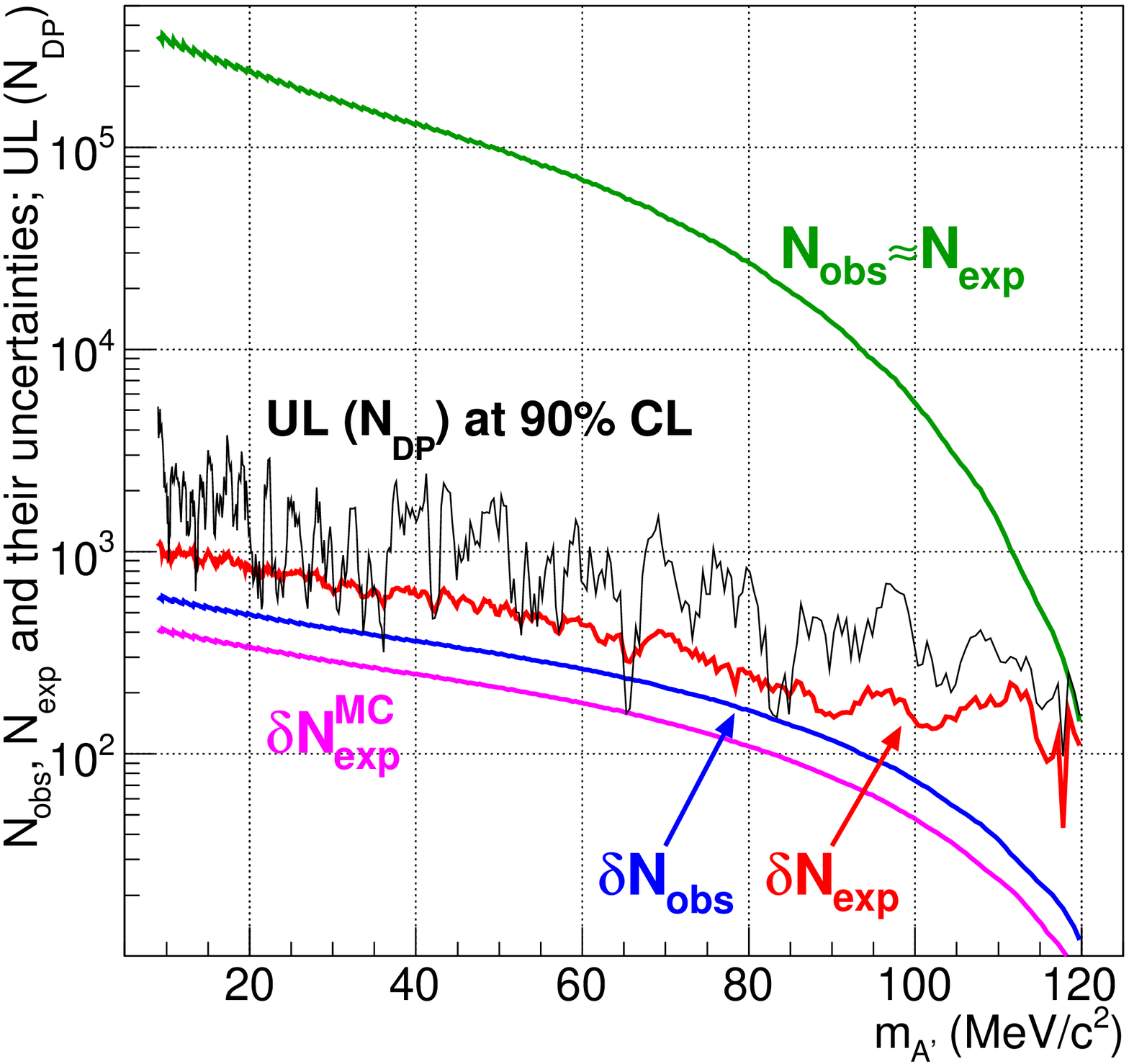}}%
\resizebox{0.5\textwidth}{!}{\includegraphics{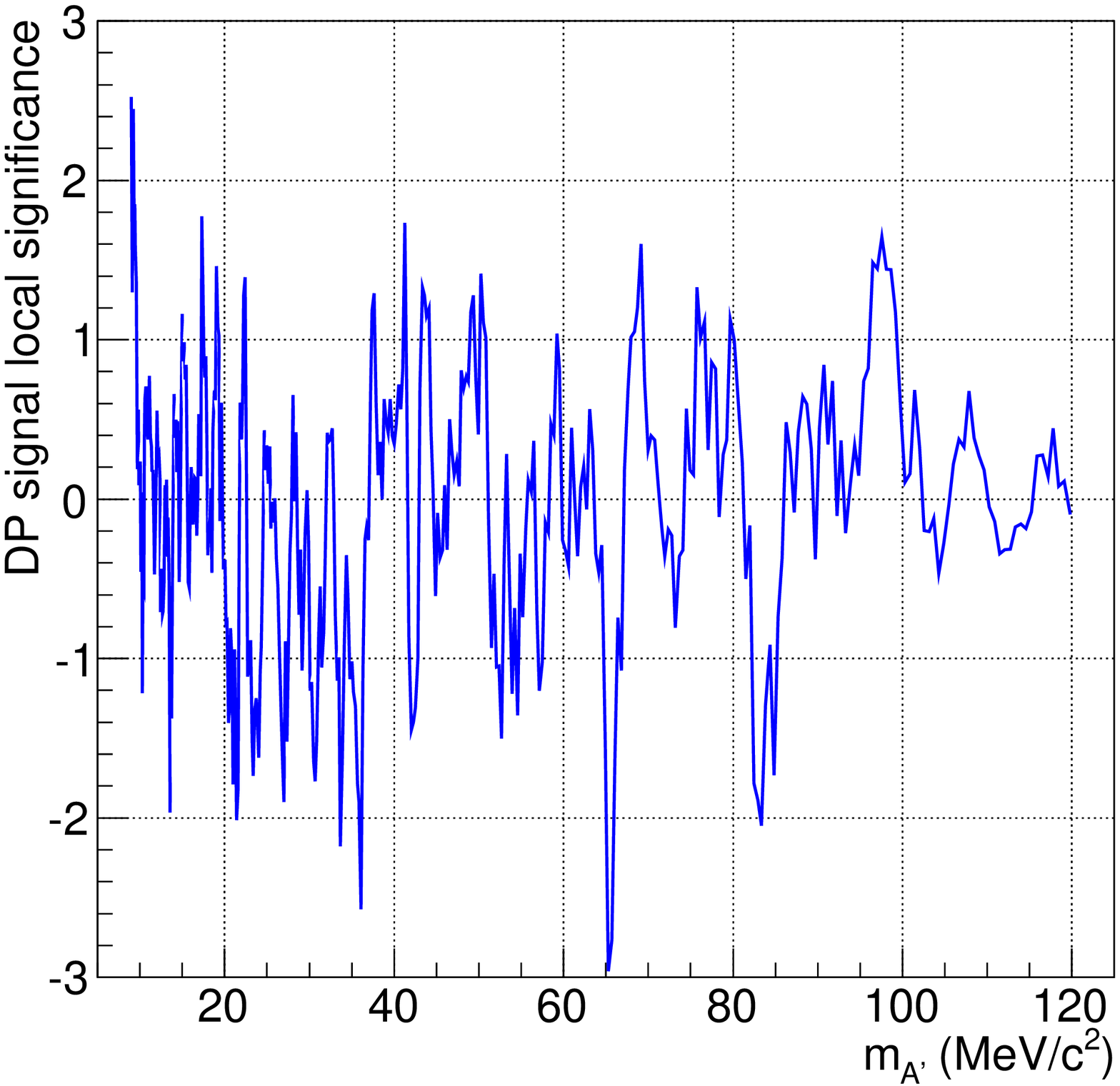}}%
\put(-247,197){\Large\bf a}
\put(-23,197){\Large\bf b}
\end{center}
\vspace{-15mm}
\caption{a)~Numbers of observed data events ($N_{\rm obs}$) and expected $\pi^0_D$ background events ($N_{\rm exp}$) passing the joint DP selection (indistinguishable in a logarithmic scale), estimated uncertainties $\delta N_{\rm obs}=\sqrt{N_{\rm exp}}$ and $\delta N_{\rm exp}$, and obtained upper limits at 90\% CL on the numbers of DP candidates ($N_{\rm DP}$) for each DP mass value $m_{A'}$. The contribution to $\delta N_{\rm exp}$ from the MC statistical uncertainty is shown separately ($\delta N_{\exp}^{\rm MC}$). The remaining and dominant component is due to the statistical errors on the trigger efficiencies measured in the DP signal region. b)~Estimated local significance of the DP signal for each $A'$ mass value. All presented quantities are strongly correlated for neighbouring DP masses as the mass step of the scan is about 6 times smaller than the signal window width.}
\label{fig:observed}
\end{figure}

For each considered DP mass value, the number of observed data events $N_{\rm obs}$ passing the joint DP selection is compared to the expected number of background events $N_{\rm exp}$. The latter is evaluated from MC simulations (Section~\ref{sec:background}), corrected for the trigger efficiency measured from control data samples passing the joint DP selection. The numbers of observed and expected events for each DP mass value and their estimated uncertainties $\delta N_{\rm obs}$ and $\delta N_{\rm exp}$ are shown in Fig.~\ref{fig:observed}a. The quantities $N_{\rm obs}$ and $N_{\rm exp}$ decrease with the assumed DP mass value due to the steeply falling $\pi^0_D$ differential decay rate (Eq.~\ref{eq:dgdxdy}) and decreasing acceptances, although the search window width increases approximately proportionally to $m_{A'}$. The uncertainty $\delta N_{\rm obs}=\sqrt{N_{\rm exp}}$ is statistical, while the uncertainty $\delta N_{\rm exp}$ has contributions from the limited size of the generated MC samples and the statistical errors on the trigger efficiencies measured in the DP signal region.

\begin{figure}[p]
\begin{center}
\resizebox{0.5\textwidth}{!}{\includegraphics{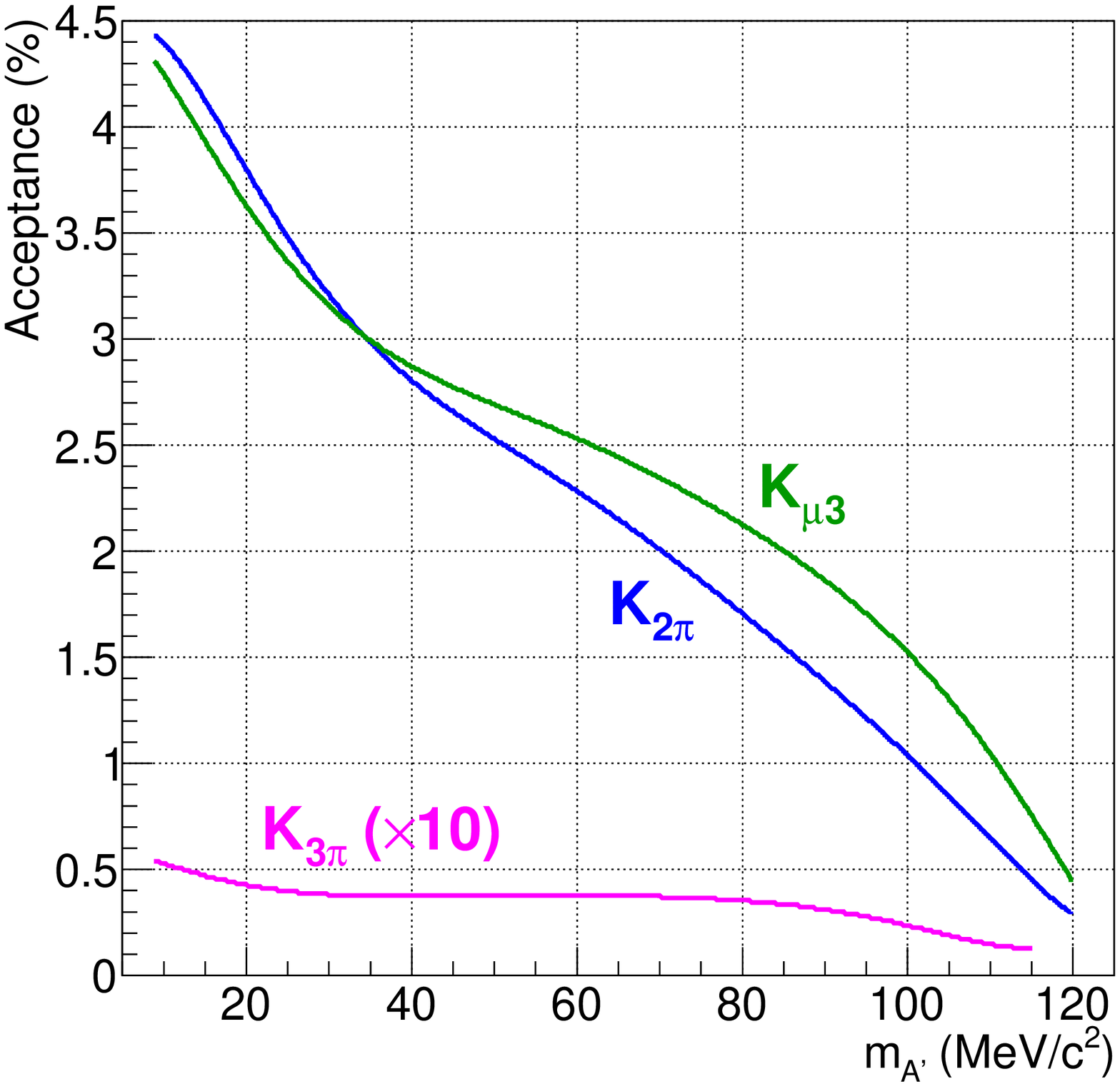}}%
\resizebox{0.5\textwidth}{!}{\includegraphics{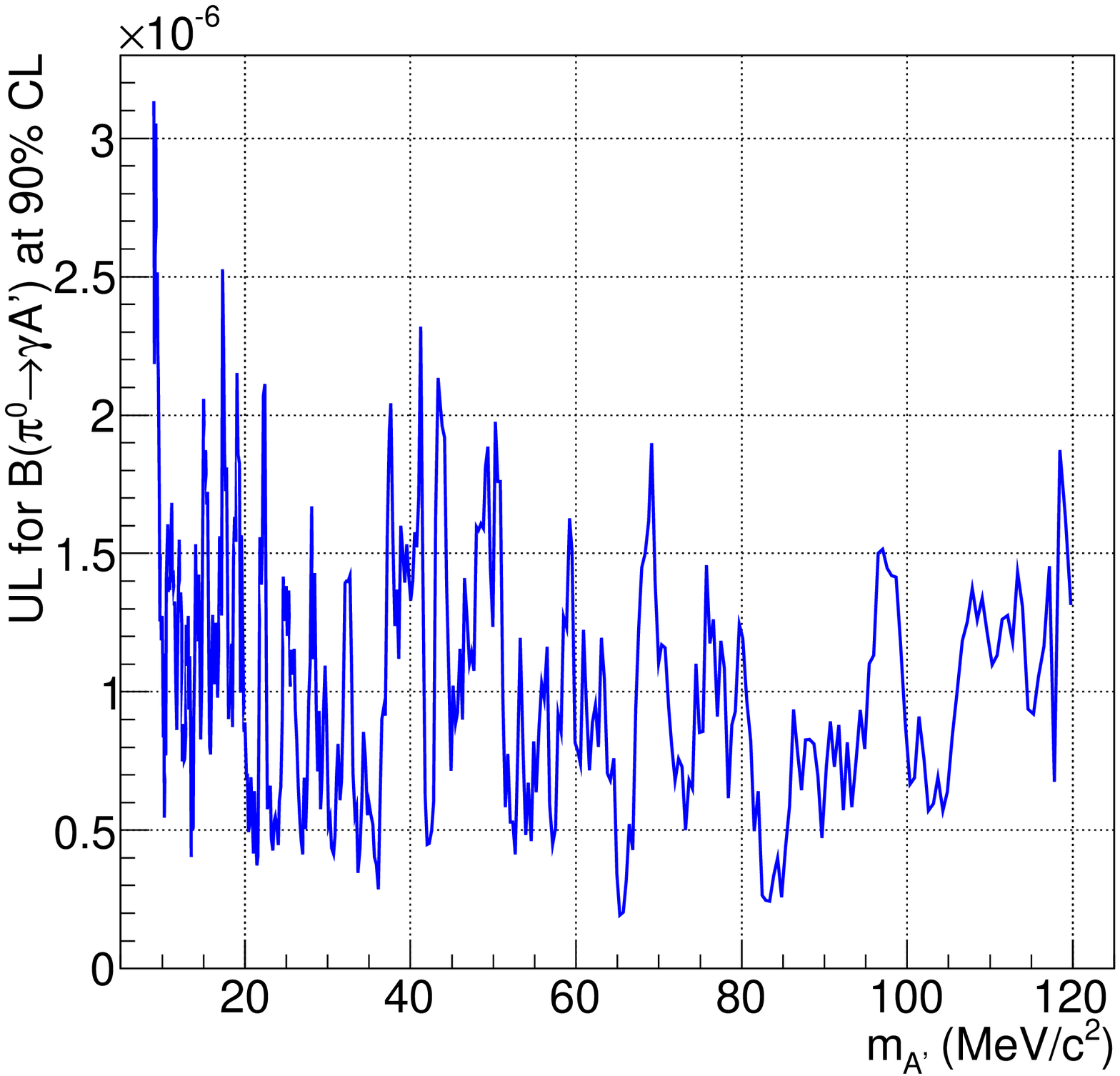}}
\put(-249,194){\Large\bf a}
\put(-21,194){\Large\bf b}
\end{center}
\vspace{-15mm}
\caption{a)~Acceptances of the joint DP selection for $K_{2\pi}$, $K_{\mu 3}$ and $K_{3\pi}$ decays followed by the prompt decay chain $\pi^0\to\gamma A'$, $A'\to e^+e^-$ depending on the assumed DP mass, evaluated with MC simulations. The $K_{3\pi}$ acceptance is scaled by a factor of 10 for visibility. b)~Obtained upper limits on ${\cal B}(\pi^0\to\gamma A')$ at 90\% CL for each DP mass value $m_{A'}$.}
\label{fig:acc-br}
\end{figure}

The local statistical significance of the DP signal for each mass value estimated as
\begin{equation}
\label{eq:significance}
Z=(N_{\rm obs}-N_{\rm exp})/\sqrt{(\delta N_{\rm obs})^2+(\delta N_{\rm exp})^2}
\end{equation}
is shown in Fig.~\ref{fig:observed}b. The local significance never exceeds $3\sigma$, therefore no DP signal is observed. Confidence intervals at 90\% CL for the number of $A'\to e^+e^-$ decay candidates for each DP mass value ($N_{\rm DP}$) are computed from $N_{\rm obs}$, $N_{\rm exp}$ and $\delta N_{\rm exp}$ using the frequentist Rolke--L\'opez method~\cite{ro01}.
The obtained upper limits on $N_{\rm DP}$ at 90\% CL are displayed in Fig.~\ref{fig:observed}a. The observed spikes in the upper limits versus the DP mass are due to the finite step of the mass scan.

Upper limits at 90\% CL on the branching fraction ${\cal B}(\pi^0\to\gamma A')$ for each DP mass value with the assumption ${\cal B}(A'\to e^+e^-)=1$ (which is a good approximation for $m_A'<2m_\mu$ if $A'$ decays to SM fermions only) are computed using the relation
\begin{equation}
\label{eq:brdp}
{\cal B}(\pi^0\to\gamma A') = \frac{N_{\rm DP}}{N_K e_1 e_2
[{\cal B}(K_{2\pi}) A_{\rm DP}(K_{2\pi}) + {\cal B}(K_{\mu 3}) A_{\rm DP}(K_{\mu 3}) +
2 {\cal B}(K_{3\pi})A_{\rm DP}(K_{3\pi})]},
\end{equation}
where $A_{\rm DP}(K_{2\pi})$, $A_{\rm DP}(K_{\mu 3})$ and $A_{\rm DP}(K_{3\pi})$ are the acceptances of the joint DP selection for $K_{2\pi}$, $K_{\mu 3}$ and $K_{3\pi}$ decays, respectively, followed by the prompt $\pi^0\to\gamma A'$, $A'\to e^+e^-$ decay chain. The trigger efficiencies $e_1$ and $e_2$ (Section~\ref{sec:flux}) are taken into account neglecting their variations over the $m_{ee}$ mass, variations measured to be at the level of a few permille.

Distributions of the angle between the $e^+$ momentum in the $e^+e^-$ rest frame and the $e^+e^-$ momentum in the $\pi^0$ rest frame are identical for the decay chain involving the DP ($\pi^0\to\gamma A'$, $A'\to e^+e^-$) and the $\pi^0_D$ decay, up to the radiative corrections relevant in the latter case but not in the former case. Therefore the acceptances for each DP mass value are evaluated with MC samples of $\pi^0_D$ decays simulated without radiative corrections. Applying radiative corrections induces a relative change of about $1\%$ for the $\pi^0_D$ acceptance. The DP acceptance dependence on the assumed DP mass is shown in Fig.~\ref{fig:acc-br}a. The second (third) term in the denominator of Eq.~\ref{eq:brdp} is typically about 20\% (less than 1\%) of the first term. The resulting upper limits on ${\cal B}(\pi^0\to\gamma A')$ are shown in Fig.~\ref{fig:acc-br}b. They are ${\cal O}(10^{-6})$ and do not exhibit a strong dependence on the DP mass, as the mass dependences of $\pi^0_D$ background level (Fig.~\ref{fig:mass}) and signal acceptances (Fig.~\ref{fig:acc-br}a) largely compensate each other.

\begin{figure}[t]
\begin{center}
\resizebox{0.55\textwidth}{!}{\includegraphics{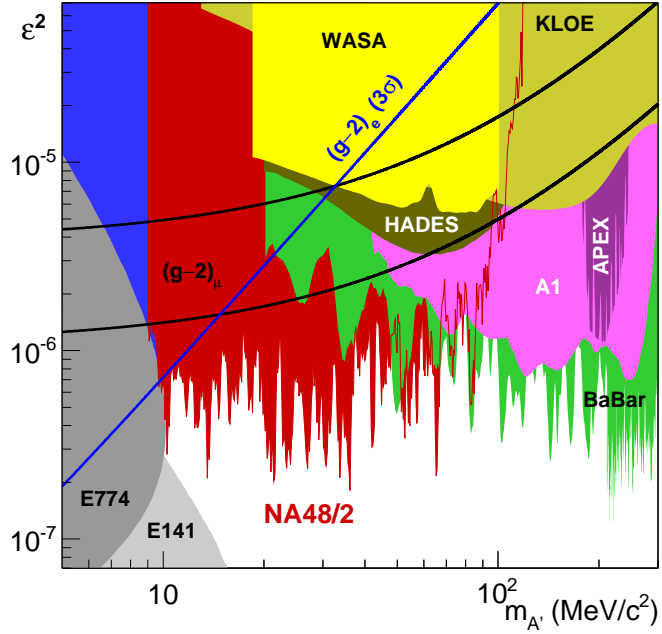}}
\end{center}
\vspace{-15mm}
\caption{Obtained upper limits at 90\% CL on the mixing parameter $\varepsilon^2$ versus the DP mass
$m_{A'}$, compared to other published exclusion limits from meson decay, beam dump and $e^+e^-$ collider experiments~\cite{an12,bab13,ad13,ag14,me14,ab11,le14}. Also shown is the band where the inconsistency of theoretical and experimental values of muon $(g-2)$ reduces to less than 2 standard deviations, as well as the region excluded by the electron $(g-2)$ measurement~\cite{po09,en12,da14}.}
\label{fig:world}
\end{figure}

Upper limits at 90\% CL on the mixing parameter $\varepsilon^2$ for each DP mass value calculated from the ${\cal B}(\pi^0\to\gamma A')$ upper limits using Eq.~\ref{eq:br} are shown in Fig.~\ref{fig:world}, together with the constraints from the SLAC E141 and FNAL E774~\cite{an12}, KLOE~\cite{bab13}, WASA~\cite{ad13}, HADES~\cite{ag14}, A1~\cite{me14}, APEX~\cite{ab11} and BaBar~\cite{le14} experiments. Also shown is the band in the ($m_{A'}$, $\varepsilon^2$) plane where the discrepancy between the measured and calculated muon $(g-2)$ values falls into the $\pm2\sigma$ range due to the DP contribution, as well as the region excluded by the electron $(g-2)$ measurement~\cite{po09,en12,da14}.

The most stringent limits on $\varepsilon^2$ obtained occur at low DP mass where the kinematic suppression of the $\pi^0\to\gamma A'$ decay is weak. The prompt DP decay assumption that is fundamental to the analysis reported here is justified a posteriori by the achieved limits. Given the 60~GeV/$c$ beam, the maximum DP mean path in the laboratory reference frame corresponds to an energy of approximately $E_{\rm max}=50~{\rm GeV}$:
\begin{equation}
L_{\rm max} \approx \frac{E_{\rm max}}{m_{A'}c^2} \cdot c\tau_{A'} \approx 0.4~{\rm mm} \times \left(\frac{10^{-6}}{\varepsilon^2}\right) \times \left(\frac{100~{\rm MeV}/c^2}{m_{A'}}\right)^2.
\end{equation}
The lowest obtained limit $\varepsilon^2 m_{A'}^2=3\times 10^{-5}~{\rm MeV}^2/c^4$ translates into a maximum DP mean path of $L_{\rm max}\approx 10~{\rm cm}$. The corresponding loss of the 3-track trigger and event reconstruction efficiency is negligible, as the offline resolution on the longitudinal coordinate of a 3-track vertex is about 1~m.

The sensitivity of the prompt $A'$ decay search is limited by the irreducible $\pi^0_D$ background. In particular, the upper limits on ${\cal B}(\pi^0\to\gamma A')$ and $\varepsilon^2$ obtained in this analysis are two to three orders of magnitude above the single event sensitivity, as seen from the upper limits on $N_{\rm DP}$ in Fig.~\ref{fig:observed}a. The achievable upper limit on $\varepsilon^2$ scales as the inverse square root of the integrated beam flux, which means that the possible improvements to be made with this technique using larger future $K^\pm$ samples are modest.

\boldmath
\section{Dark photon search in the $K^\pm\to\pi^\pm A'$ decay}
\unboldmath

An alternative way to search for the DP in $K^\pm$ decays is via the $K^\pm\to\pi^\pm A'$ decay followed by the prompt $A'\to\ell^+\ell^-$ decay ($\ell=e,\mu$). This decay chain provides sensitivity to the DP in the mass range $2m_e<m_{A'}<m_K-m_\pi$. The expected branching fraction value is ${\cal B}(K^\pm\to\pi^\pm A') < 2\cdot 10^{-4}\varepsilon^2$ over the whole allowed $m_{A'}$ range~\cite{da14}, in contrast to ${\cal B}(\pi^0\to\gamma A') \sim \varepsilon^2$ for $m_{A'}<100~{\rm MeV}/c^2$. In the NA48/2 data sample, the suppression of the DP production in the $K^+$ decay with respect to its production in the $\pi^0$ decay is partly compensated by the favourable $K^\pm$/$\pi^0$ production ratio, lower background (mainly from $K^\pm\to\pi^\pm\ell^+\ell^-$ for $\ell=\mu$ or $m_{A'}>m_{\pi^0}$) and higher acceptance~\cite{ba09,ba11}.

For the $A'\to e^+e^-$ decay, the expected sensitivity of the NA48/2 data sample to $\varepsilon^2$ is maximum in the mass interval $140~{\rm MeV}/c^2<m_{A'}<2m_\mu$, where the $K^\pm\to\pi^\pm A'$ decay is not kinematically suppressed, the $\pi^0_D$ background is absent, and ${\cal B}(A'\to e^+e^-)\approx 1$ assuming that the DP decays only into SM fermions. In this $m_{A'}$ interval, the expected NA48/2 upper limits have been computed to be in the range $\varepsilon^2=(0.8-1.1)\times 10^{-5}$ at 90\% CL, in agreement with earlier generic estimates~\cite{po09,da14}. This sensitivity is not competitive with the existing exclusion limits.

\section*{Conclusions}

A search for the dark photon (DP) production in the $\pi^0\to\gamma A'$ decay followed by the prompt $A'\to e^+e^-$ decay has been performed using the data sample collected by the NA48/2 experiment in 2003--2004. No DP signal is observed, providing new and more stringent upper limits on the mixing parameter $\varepsilon^2$ in the mass range 9--70 MeV/$c^2$. In combination with other experimental searches, this result rules out the DP as an explanation for the muon $(g-2)$ measurement under the assumption that the DP couples to quarks and decays predominantly to SM fermions. The NA48/2 sensitivity to the dark photon production in the $K^\pm\to\pi^\pm A'$ decay has also been evaluated.


\section*{Acknowledgements}

We express our gratitude to the staff of the CERN laboratory and the technical staff of the participating universities and laboratories for their efforts in the operation of the experiment and data processing. We thank Bertrand Echenard, Tom\'a\v s Husek, Karol Kampf, Michal Koval, Nicolas Lurkin, Karim Massri,
Ji\v r\'i Novotn\'y and Tommaso Spadaro for their contributions.




\begin{thebibliography}{99}
%
\bibitem{ho86}
B. Holdom, Phys. Lett. {\bf B166} (1986) 196.
%
\bibitem{po09}
M.~Pospelov, Phys. Rev. {\bf D80} (2009) 095002.
%
\bibitem{batell09}
B. Batell, M. Pospelov and A. Ritz, Phys. Rev. {\bf D80} (2009) 095024.
%
\bibitem{fa07}
V. Fanti {\it et al.} (NA48 collaboration), Nucl. Instrum. Methods {\bf A574} (2007) 433.
%
\bibitem{ba07}
J.R.~Batley {\it et al.} (NA48/2 collaboration), Eur. Phys. J. {\bf C52} (2007) 875.
%
\bibitem{geant}
GEANT detector description and simulation tool,\\CERN program library
long writeup W5013 (1994).
%
\bibitem{ga06}
C. Gatti, Eur. Phys. J. {\bf C45} (2006) 417.
%
\bibitem{mi72}
K.O.~Mikaelian and J.~Smith, Phys. Rev. {\bf D5} (1972) 1763.
%
\bibitem{hu15}
T.~Husek, K.~Kampf and J.~Novotn\'y, arXiv:1504.06178.
%
\bibitem{la85}
L.G. Landsberg, Phys. Rept. {\bf 128} (1985) 301.
%
\bibitem{ho14}
M. Hoferichter {\it et al.}, Eur. Phys. J. {\bf C74} (2014) 3180.
%
\bibitem{pdg}
K.A. Olive {\it et al.} (Particle Data Group), Chin. Phys. {\bf C38} (2014) 090001.
%
\bibitem{cello}
H.-J.~Behrend {\it et al.} (CELLO collaboration), Z. Phys. {\bf C49} (1991) 401.
%
%
%
\bibitem{ka06}
K.~Kampf, M.~Knecht and J.~Novotn\'y, Eur. Phys. J. {\bf C46} (2006) 191.
%
\bibitem{ro01}
W.A.~Rolke and A.M.~L\'opez, Nucl. Instrum. Meth. {\bf A458} (2001) 745.
%
\bibitem{an12}
S.~Andreas, C.~Niebuhr and A.~Ringwald, Phys. Rev. {\bf D86} (2012) 095019.
%
\bibitem{bab13}
D.~Babusci {\it et al.} (KLOE-2 collaboration), Phys. Lett. {\bf B720} (2013) 111.
%
\bibitem{ad13}
P.~Adlarson {\it et al.} (WASA-at-COSY collaboration), Phys. Lett. {\bf B726} (2013) 187.
%
\bibitem{ag14}
G.~Agakishev {\it et al.} (HADES collaboration), Phys. Lett. {\bf B731} (2014) 265.
%
\bibitem{me14}
H.~Merkel {\it et al.} (A1 collaboration), Phys. Rev. Lett. {\bf 112} (2014) 221802.
%
\bibitem{ab11}
S.~Abrahamyan {\it et al.} (APEX collaboration), Phys. Rev. Lett. {\bf 107} (2011) 191804.
%
\bibitem{le14}
J.P.~Lees {\it et al.} (BaBar collaboration), Phys. Rev. Lett. {\bf 113} (2014) 201801.
%
\bibitem{en12}
M.~Endo, K.~Hamaguchi and G.~Mishima, Phys. Rev. {\bf D86} (2012) 095029.

\bibitem{da14}
H.~Davoudiasl, H.-S.~Lee and W.J.~Marciano, Phys. Rev. {\bf D89} (2014) 095006.
%
\bibitem{ba09}
J.R.~Batley {\it et al.} (NA48/2 collaboration), Phys. Lett. {\bf B677} (2009) 246.
%
\bibitem{ba11}
J.R.~Batley {\it et al.} (NA48/2 collaboration), Phys. Lett. {\bf B697} (2011) 107.

\end{thebibliography}
\end{document}